\documentclass[aps,prl,reprint,preprintnumbers,groupedaddress] {revtex4-1}
\input{Qcircuit}
\usepackage{slashbox}
\usepackage{graphicx}
\usepackage{amsmath}
\usepackage{amsfonts}
\usepackage{amssymb}
\usepackage{amstext}
\usepackage{textcomp}
\usepackage[makeroom]{cancel}
\usepackage{hyperref}
\usepackage{braket}
\usepackage{natbib}
 \def\be{\begin{equation}}
 \def\ee{\end{equation}}
 \def\bes{\begin{eqnarray}}
 \def\ees{\end{eqnarray}}
\begin{document}
\title
	{Quantum Prisoners' Dilemma under Enhanced Interrogation}
\author{George Siopsis} \email{siopsis@tennessee.edu}
\affiliation{Department of Physics and Astronomy,
		The University of Tennessee, Knoxville, TN 37996-1200, USA}
\author{
	Radhakrishnan Balu} \email{radhakrishnan.balu.civ@mail.mil}
\affiliation{
	Computer and Information Sciences Directorate, Army Research Laboratory, Adelphi, MD, 21005-5069, USA. }

\author{Neal Solmeyer}\email{neal.e.solmeyer.civ@mail.mil} 
\affiliation{
	Sensors and Electron Devices Directorate, Army Research Laboratory, Adelphi, MD, 21005-5069, USA. }
\date{\today}
\begin{abstract}
	In the quantum version of prisoners' dilemma, each prisoner is equipped with a single qubit that the interrogator can entangle. We enlarge the available Hilbert space by introducing a third qubit that the interrogator can entangle with the other two. We discuss an enhanced interrogation technique based on tripartite entanglement and analyze Nash equilibria. We show that for tripartite entanglement approaching a W-state, there exist Nash equilibria that coincide with the Pareto optimal choice where both prisoners cooperate. Upon continuous variation between a W-state and a pure bipartite entangled state, the game is shown to have a surprisingly rich structure. The role of bipartite and tripartite entanglement is explored to explain that structure. 
\end{abstract}
\maketitle

Quantum games as a field received a lot of attention from the early works of Meyer \cite{Meyer1999}, and has grown steadily ever since. Connections exist between quantum games and various other fields, such as Bell non-locality \cite{Brunner2013} and quantum logic \cite{Piotrowski2003}, to name a few. Various aspects of quantum games, including the role of entanglement and multiple player extensions, explored by different authors can be found in references \cite{Flitney2005,Iqbal2002,Hayden2002}. A solution to the quantum prisoners' dilemma in which players have a Nash equilibrium that is Pareto optimal ignited interest in quantum games \cite{Eisert1999}. However, this initial formulation drew criticism because it dramatically restricted the strategy space of the players and did not persist under maximal entanglement if arbitrary quantum strategies were allowed \cite{Benjamin2001}.

In this work, we enlarge the Hilbert space in a minimal way in order to arrive at a Nash equilibrium (NE) that is Pareto-optimal with maximal entanglement. In addition to the two player qubits, we consider a resource qubit that the referee, or interrogator, has access to. In the three-qubit Hilbert space, one can introduce tripartite entanglement which will lead to a much richer Nash equilibrium structure.  Interestingly, for tripartite entanglement close to maximum (approaching a W-state), we obtain Nash equilibria that coincide with Pareto optimal points. The bipartite entanglement between the players' qubits partially explains the structure. In addition, it is shown that there is no NE for a GHZ state, which has a fundamentally different type of entanglement \cite{bibdur}.

We briefly review the standard formulation of the prisoners' dilemma. It is a game of two players, Alice and Bob, who must decide independently whether they defect (strategy $D$) or cooperate (strategy $C$). Each player receives a payoff and chooses a strategy that maximizes it. For concreteness, we shall use Table \ref{table:1} to determine the payoff \cite{bibu}.
\begin{table}[ht!]
\begin{tabular}{|c||c|c|}\hline
\backslashbox{\text{Alice}}{\text{Bob}} &\ \ \ \ \ \ \ \ \ C \ \ \ \ \ \ \ \ \ & \ \ \ \ \ \ \ \ \ D \ \ \ \ \ \ \ \ \ \\
\hline\hline	C & (11,9) & (1,10) \\\hline
	D & (10,1) & (6,6) \\\hline
\end{tabular}
\caption{\label{table:1}Prisoners' payoff matrix. In each pair, the first (second) entry is payoff for Alice (Bob).}
\end{table}
The best strategy is $CC$ (both prisoners cooperate), which is the Pareto optimal choice. However, by making unilateral decisions, they choose $DD$, which is the Nash equilibrium.

The classical game outlined above has been quantized as follows. Suppose that Alice and Bob are in possession of one qubit each with the state $|0\rangle$ ($|1\rangle$) corresponding to the choice $C$ ($D$). Thus, the set of four classical possibilities $\{ CC, CD, DC, DD \}$ corresponds to the basis $\{ |00\rangle, |01\rangle, |10\rangle, |11\rangle  \}$. A general quantum state is a linear superposition of the four basis vectors,
\be \label{eqpsi}|\psi\rangle = \sum_{x=0}^3 a_x |x\rangle \ee
with $x$ written in binary notation. The payoff for Alice is then
\be\label{eqpay} \$_A = \sum_{x=0}^{3} \$_{A,x} |a_x|^2 \ee
where $\$_{A,x}$ are the classical payoff values for Alice given in Table \ref{table:1}, and similarly for Bob.

    \begin{figure}[ht!]
	\centering
	\includegraphics[scale=.4]{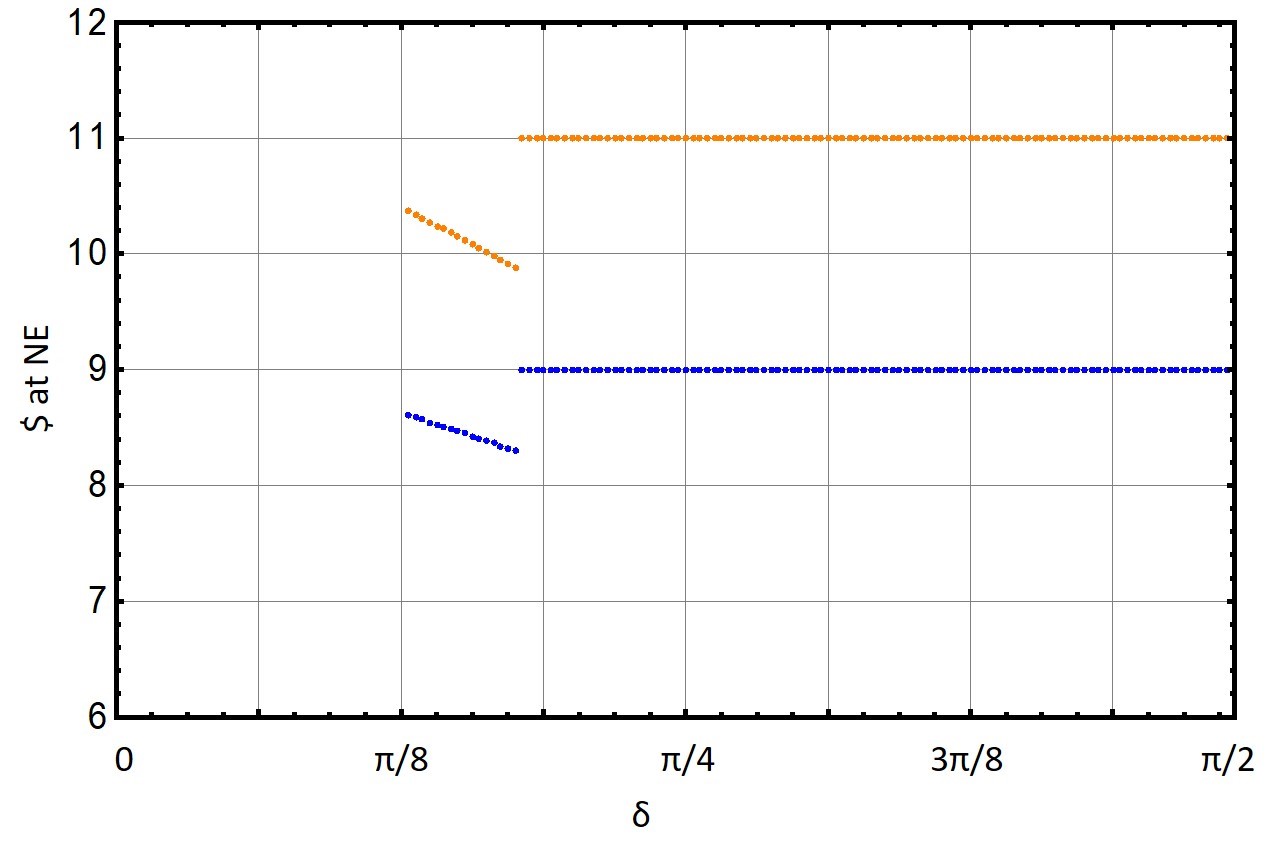}
	\caption{\label{fig:1} Payoffs for $A$ (upper line) and $B$ (lower line) \emph{vs}.\ $\delta$ for $\gamma = \pi/2$. The strategies correspond to (C,C), or $U_A=U_B = I$.}
\end{figure}

Alice and Bob start with qubits in the state $|0\rangle$. The interrogator then entangles them by applying the unitary
\be\label{eq1} J(\gamma) = e^{i\frac{\gamma}{2} \sigma_X\otimes \sigma_X} = \cos\frac{\gamma}{2} + i \sigma_X\otimes \sigma_X \sin\frac{\gamma}{2} \ee
Then the prisoners choose strategies represented by applying unitaries $U_A$ and $U_B$, respectively, to their qubits. We adopt the use of quaternionic strategy choices \cite{Landsburg}, i.e., the Pauli matrices and the identity matrix, so that there are 4 strategy choices, $\{I, \sigma_X, \sigma_Y, \sigma_Z\}$. This choice of a strategy set ensures that the analysis of the behavior of the game captures the behavior as if the players had complete freedom to play any quantum strategy. 

Finally, the interrogator reverses the setup, by applying the unitary $J^\dagger (\gamma) = J(-\gamma)$ to the prisoners' qubits. The circuit is shown below:
\be 
\Qcircuit @C=1em @R=.7em {
		\lstick{A\ \ \ \ket{0}}  &\multigate{1}{J(\gamma)}  &\gate{U_A}  &\multigate{1}{J^\dagger (\gamma)}  &\meter & \qw \\
	\lstick{B\ \ \ \ket{0}}  &\ghost{J(\gamma)} &\gate{U_{B}}  &\ghost{J^\dagger (\gamma)}  &\meter & \qw \\
}
\label{eq:circ0}\ee
The final state is of the form \eqref{eqpsi} for which payoffs $\$_{A,B}$ are computed using \eqref{eqpay}. This results in new Nash equilibria which are distinct from the classical one. They reduce to the classical case if $\gamma = 0$ (no entanglement). Nash equilibria exist for degree of entanglement $\gamma$ below a certain value ($\gamma < 1.15$). In the case of maximal entanglement, no Nash equilibria exist \cite{bibu}.

Our enlarged Hilbert space consists of a resource qubit $Q$, as well qubits $A$ and $B$ that belong to the two prisoners, respectively. The prisoners' payoffs are given in Table \ref{table:1}. Each qubit is in the state $|0\rangle$ initially.
The interrogator starts by entangling $A$ and $B$ using the unitary \eqref{eq1}, as before.
Subsequently, he uses $A$ as control to apply the rotation $e^{i\delta \sigma_X}$ on qubit $Q$, with a control parameter $\delta$ that can vary. The state of the system $QAB$ becomes
\be |\phi(\gamma,\delta)\rangle = \cos\frac{\gamma}{2} |000\rangle + i\sin\frac{\gamma}{2}  (\cos\delta |0\rangle + i\sin\delta |1\rangle )|11\rangle\ee
Then he uses $B$ and $Q$ as controls to flip $A$. The state becomes
\be\label{eq6} |\phi(\gamma,\delta)\rangle = \cos\frac{\gamma}{2} |000\rangle + i\sin\frac{\gamma}{2} \cos\delta |011\rangle -\sin\frac{\gamma}{2}\sin\delta |101\rangle \ee
Notice that, if $\tan\frac{\gamma}{2} = \sqrt{2}$, and $\delta = \frac{\pi}{4}$,  it is maximally entangled. We obtain
\be\label{eq6W} |\Phi\rangle = \frac{1}{\sqrt{3}} \left( |000\rangle + i |011\rangle - |101\rangle \right) \ee
This state is the tripartite $W$-state \cite{bibdur},
\be\label{eqW} |W\rangle = \frac{1}{\sqrt{3}} \left( |001\rangle +  |010\rangle + |100\rangle \right) \ee
up to single-qubit transformations 
(application of $\sigma_X$ on the third qubit (prisoner B), and phase changes).

If $\delta =0$, this is equivalent to the setup considered above.

Then the prisoners apply strategies $U_A$ and $U_{B}$, respectively. After they are done strategizing, the referee reverses the setup of the enhanced interrogation.

The circuit is shown below:
\be \ \ \ \ \ \ \ 
\Qcircuit @C=.2em @R=.7em {
	\lstick{Q\ \ \ \ket{0}}   &\qw &\qw &\gate{e^{i\delta \sigma_X}} &\ctrl{1} &\qw &\ctrl{1} &\gate{e^{-i\delta \sigma_X}} &\qw  &\qw &\qw &\qw \\
	\lstick{A\ \ \ \ket{0}}  &\qw &\multigate{1}{J(\gamma)} &\ctrl{-1} &\targ &\gate{U_A} &\targ &\ctrl{-1} &\multigate{1}{J^\dagger (\gamma)}  &\qw &\meter &\qw \\
\lstick{B\ \ \ \ket{0}}  &\qw &\ghost{J(\gamma)} &\qw &\ctrl{-1} &\gate{U_{B}} &\ctrl{-1} &\qw &\ghost{J^\dagger (\gamma)}  &\qw &\meter &\qw \\
}
\label{eq:circ1}\ee
For $\gamma > 1.15$, there are no NE with the circuit \eqref{eq:circ0}. In fig.\ \ref{fig:1}, we show the results for the modified circuit, \eqref{eq:circ1} for maximal entanglement, i.e.  $\gamma = \frac{\pi}{2}$. The payoff for both players at the NE is plotted as a function of the control parameter, $\delta$. For $\delta < 0.41$, there are no NE. But for $\delta > 0.56$, they exist. In fact, they coincide with the Pareto optimal choice where both prisoners confess! This equilibrium is formed by both players playing the identity matrix, i.e., $I \otimes I$ . 

For $\delta \in (0.41,0.56)$, a different NE exists that is not the Pareto optimal choice. This equilibrium is formed by players $A$ and $B$ playing the strategy choices $\sigma_Z \otimes \sigma_X$.

\begin{figure}[ht!]
	\centering
	\includegraphics[scale=.4]{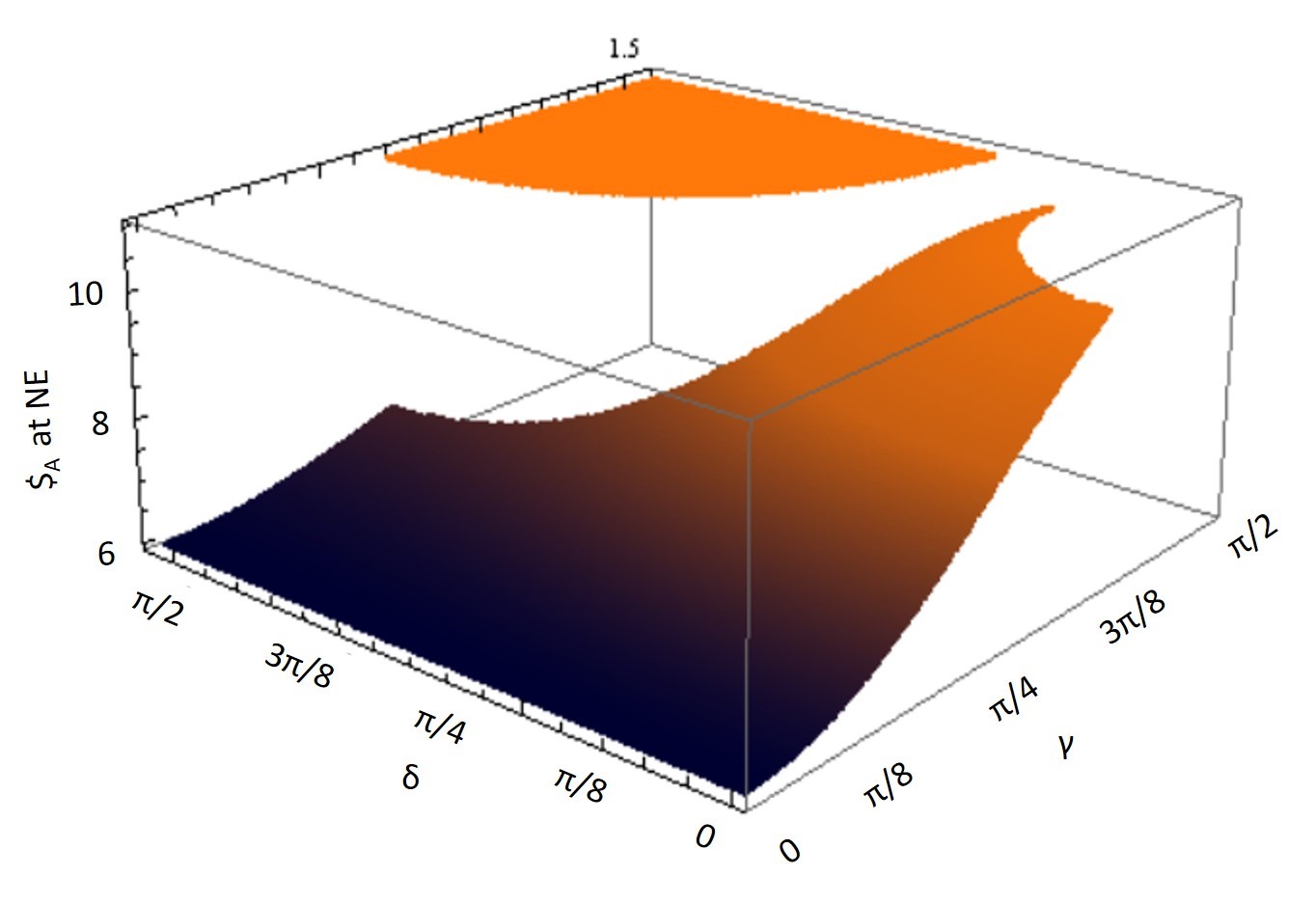}
	\caption{\label{fig:2} Payoffs for $A$ are shown for all values of $\delta$ and $\gamma$. The payoffs for $B$ are qualitatively the same, with different values, owing to the asymmetry of the payoff matrix.}
\end{figure}

In fig.\ \ref{fig:2}, we plot the results as a function of both bipartite entanglement, and the control parameter $\delta$. For $\delta = 0$, the curve reproduces the results that have been previously found for the prisioners' dilemma with partial entanglement. As $\delta$ increases, the payoff begins to lower generally, and the region near maximal entanglement where there is no NE shrinks. Where there are multiple NE, we plot the payoff for the NE with the highest payoff. Qualitatively, there is no NE when the control and entanglement parameters fall within an ellipse centered on  $\delta = 0, \gamma = \pi/2$ with radii $r_\gamma = 0.42$ and $r_\delta = 0.41$.  In fact, this is the only region of the parameter space that has no NE.

In addition, the Pareto optimal solutions are shown to form a plateau centered on $\delta = \pi/2, \gamma = \pi/2$. Numerically, fitting the data to an ellipse, we find that when the control and entanglement parameters fall within an ellipse centered on $\delta = \pi/2, \gamma = \pi/2$, with radii $r_\gamma = 0.90$ and $r_\delta = 1.00$, the payoff at the NE is equal to the Pareto-optimal choices. Outside of that region, there is a discontinuous drop of the payoff at NE that continuously deforms into the NE that exists at $\delta = 0$ in the standard quantum prisoners' dilemma with partial entanglement.  

\begin{figure}[ht!]
	\centering
	\includegraphics[scale=.4]{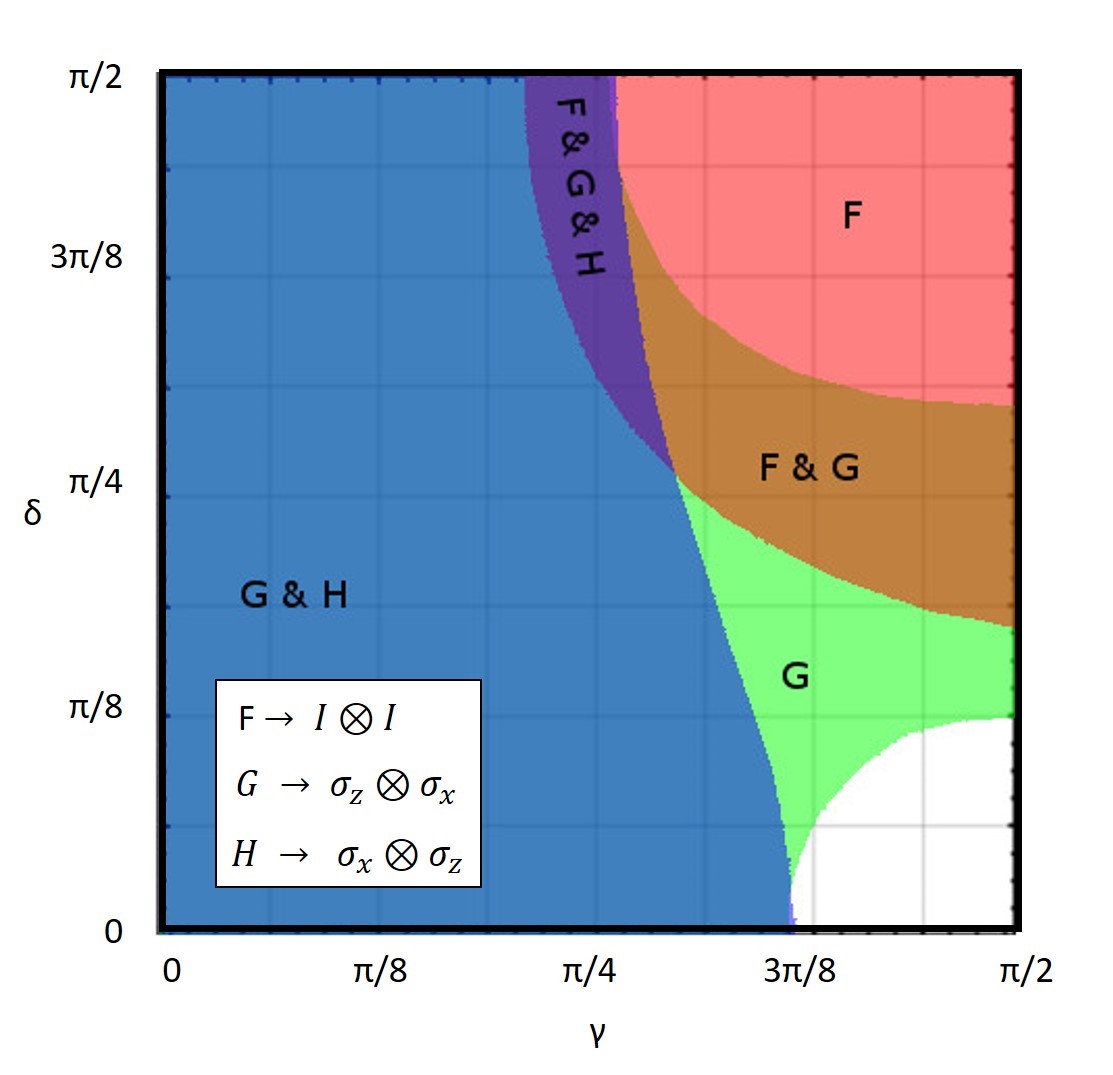}
	\caption{\label{fig:3} The NE in different regions of the plot. The various NE span different parameter regions of the plot, there is an ellipse-like region centered around $\gamma = \pi/2, \delta = 0$ that has no NE, and an elipse-like region centered around $\gamma = \pi/2, \delta = \pi/2$ which has a Pareto optimal NE.}
\end{figure}

The various types of equilibria that occur show a surprisingly complicated phase-diagram-like structure. This structure is shown in   fig.\ \ref{fig:3}. The strategy choices of the NE in the Pareto optimal plateau are $I \otimes I$ (type $F$ in fig.\ \ref{fig:3}). There are two types of equilibria that exist with partial entanglement: $\sigma_Z \otimes \sigma_X$ (type $G$) and $\sigma_X \otimes \sigma_Z$ (type $H$). Type $B$ occurs everywhere except for two elliptical regions: a region centered on $\gamma = \pi/2, \delta = \pi/2$ with radii $r_\gamma = 0.75$ and $r_\delta = 0.60$, and a region centered on $\gamma = \pi/2, \delta = 0$ with radii $r_\gamma = 0.42$ and $r_\delta = 0.41$. It should be noted that the so-called elliptical regions, in reality have a slightly different curvature than ellipses, but their theoretical and analytic description remains elusive. 

 The type $H$  only occurs when $\gamma$ is less than a curved boundary value near $\gamma \sim \frac{5 \pi}{16}$.There are regions which only have types $F$ or  $G$, and also regions which overlap with types $\{F,G \}$, or $\{G,H\}$, or  $\{F,G,H\}$.  Type $G$ and $H$ equilibria give the same payoff for the players, while type $F$ is different. Along the line of $\gamma  = 0$, there are two additional equilibria not shown on the graph. They are given by $\sigma_Z \otimes \sigma_Z$ and $\sigma_X \otimes \sigma_X$, and have the same payoff as types $G$ and $H$.

\begin{figure}
\includegraphics[width= \columnwidth]{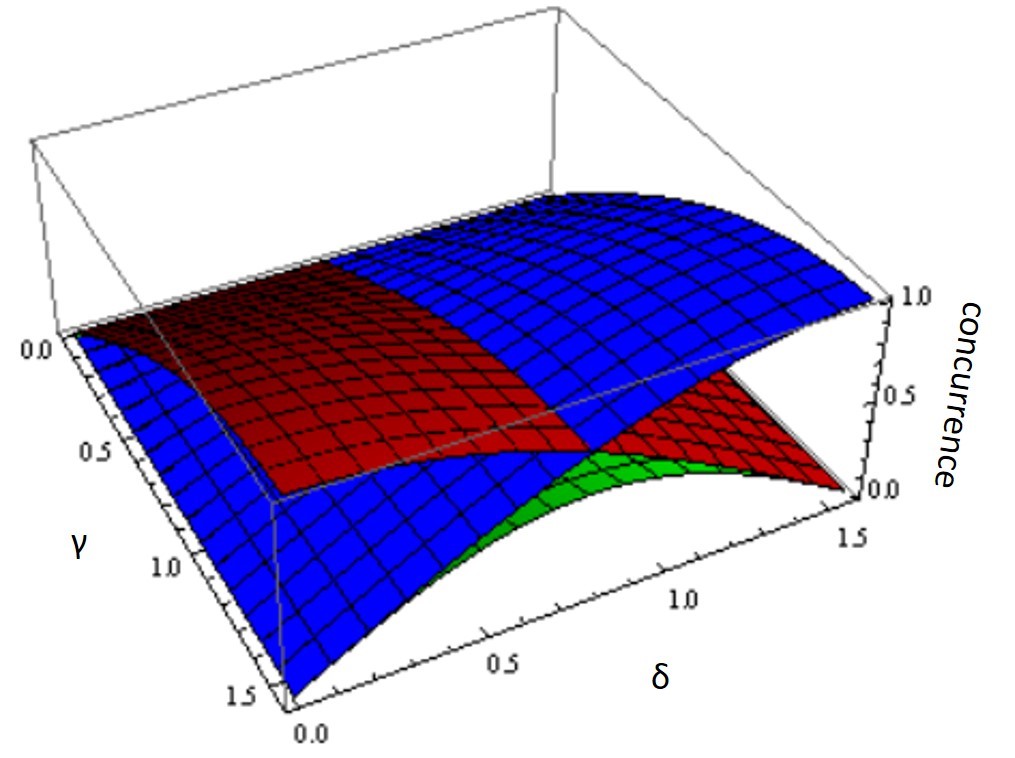}
\caption{\label{fig:conc}Bipartite entanglements -- Red: entanglement between A and B. Blue: entanglement between Q and B. Green: entanglement between Q and A.}
\end{figure}

In order to gain insight into the role of entanglement in the NE, the three bipartite entanglements are computed for the qubits just before the players apply their strategy choices. A partial trace is taken of one qubit to obtain the density matrix of the other two, then the  concurrence of the resulting density matrix is computed. The three values of concurrence, i.e., entanglement between A and B, between Q and B, and between Q and A, are computed for all values of $\delta$ and $\gamma$, and the results are shown in fig.\ \ref{fig:conc}. 

The red surface of fig.\ \ref{fig:conc} represents the bipartite entanglement between A and B showing that it does not remain maximal for $\gamma = \frac{\pi}{2}$ as $\delta$ increases. The region with no NE follows a contour of 0.91 concurrence between qubits A and B suggesting that the absence of a NE can be explained by the bipartite entanglement between A and B being larger than a threshold value, as is the case in the original quantum prisoners' dilemma. 

    \begin{figure}[ht!]
	\centering
	\includegraphics[scale=.5]{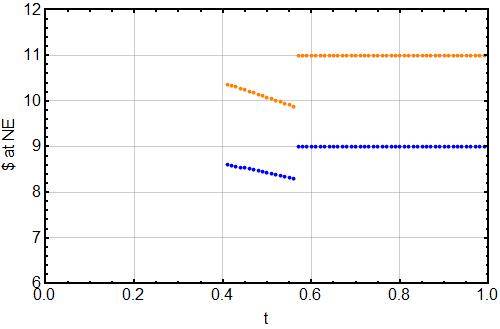}
	\caption{\label{fig:5} Payoffs for $A$ (upper line) and $B$ (lower line) \emph{vs}.\ $t$, where $\gamma = t \tan^{-1} \sqrt{2} + (1-t) \frac{\pi}{2}$, $\delta = t \frac{\pi}{4}$, and $\eta = (1-t) \frac{\pi}{2}$, interpolating between the GHZ-type state \eqref{eq6GHZ} ($t=0$) and the $W$-type state \eqref{eq6W} ($t=1$).}
\end{figure}

The region where there is no NE for type G in fig.\ \ref{fig:3} and the plateau of Pareto optimal solutions appear as if they might follow contours of the concurrence between Q and B, yet upon close analysis, this is not found to be the case, suggesting that their explanation requires an analysis of more than just the bipartite entanglement.

As noted above, for certain values of $\gamma$ and $\delta$, NE coincide with the Pareto optimal choice. This range of parameters includes the maximally entangled state \eqref{eq6} (with $\tan \frac{\gamma}{2} = \sqrt{2}$, $\delta = \frac{\pi}{4}$), which can be transformed to 
the tripartite $W$-state \eqref{eqW} 
with single-qubit transformations.
Then the question arises whether the other independent tripartite maximally entangled state (GHZ state),
\be |\text{GHZ} \rangle = \frac{1}{\sqrt{2}} \left( |000\rangle + |111\rangle \right) \ee
has a similar effect on NE. To answer this question, we draw the circuit:

\begin{widetext}
\be \ \ \ \ \ \ \ 
\Qcircuit @C=.2em @R=.7em {
	\lstick{Q\ \ \ \ket{0}}   &\qw &\qw &\gate{e^{i\delta \sigma_X}} &\ctrl{1}  &\gate{e^{i\eta \sigma_X}} &\qw &\gate{e^{-i\eta \sigma_X}} &\ctrl{1} &\gate{e^{-i\delta \sigma_X}} &\qw  &\qw &\qw &\qw \\
	\lstick{A\ \ \ \ket{0}}  &\qw &\multigate{1}{J(\gamma)} &\ctrl{-1} &\targ &\ctrl{-1} &\gate{U_A} &\ctrl{-1} &\targ &\ctrl{-1} &\multigate{1}{J^\dagger (\gamma)}  &\qw &\meter &\qw \\
	\lstick{B\ \ \ \ket{0}}  &\qw &\ghost{J(\gamma)} &\qw &\ctrl{-1} &\qw &\gate{U_{B}} &\qw &\ctrl{-1} &\qw &\ghost{J^\dagger (\gamma)}  &\qw &\meter &\qw \\
}
\label{eq:circW}\ee
\end{widetext}
In the circuit above, there is an additional step the interrogator performs before $A$ and $B$ get a chance to play. It depends on an additional parameter $\eta$. The interrogator uses qubit $A$ as control to apply the rotation $e^{i\eta \sigma_X}$ on the resource qubit $Q$. If $\eta = 0$, this reduces to the previous setup. In general, the state \eqref{eq6} is transformed to
\bes\label{eq6a} &&\cos\frac{\gamma}{2} |000\rangle + i\sin\frac{\gamma}{2} \cos\delta \cos\eta|011\rangle \nonumber\\
&&+ i\sin\frac{\gamma}{2} \cos\delta \sin\eta|111\rangle -\sin\frac{\gamma}{2}\sin\delta |101\rangle \ees
By varying the parameters, we can interpolate between the $W$-type state \eqref{eq6W} (with $\tan\frac{\gamma}{2} = \sqrt{2}, \delta = \frac{\pi}{4}, \eta = 0$), and the GHZ-type state (with $\gamma = \frac{\pi}{2}, \delta =0, \eta = \frac{\pi}{2}$),
\be\label{eq6GHZ} |\Phi' \rangle = \frac{1}{\sqrt{2}} \left( |000\rangle + i |111\rangle \right) \ee
Notice that with the choice of parameters that yield the GHZ-type state \eqref{eq6GHZ}, the circuit \eqref{eq:circW} reduces to
\be \ \ \ \ \ \ \ 
\Qcircuit @C=.5em @R=.7em {
	\lstick{Q\ \ \ \ket{0}}   &\qw &\qw   &\gate{\sigma_X} &\qw  &\gate{\sigma_X} &\qw  &\qw &\qw &\qw \\
	\lstick{A\ \ \ \ket{0}}  &\qw &\multigate{1}{J(\frac{\pi}{2})}  &\ctrl{-1} &\gate{U_A} &\ctrl{-1} &\multigate{1}{J^\dagger (\frac{\pi}{2})}  &\qw &\meter &\qw \\
	\lstick{B\ \ \ \ket{0}}  &\qw &\ghost{J(\frac{\pi}{2})}  &\qw &\gate{U_{B}}  &\qw &\ghost{J^\dagger (\frac{\pi}{2})}  &\qw &\meter &\qw \\
}
\label{eq:circGHZ}\ee
Therefore, the state of the resource qubit $Q$ does not affect the prisoners. Without a resource qubit, it is known that the maximally entangled state does not have NE. Consequently, the GHZ-type state \eqref{eq6GHZ} yields no NE. An interpolation between the $W$-type state \eqref{eq6W} and the GHZ-type state \eqref{eq6GHZ} is shown in fig.\ \ref{fig:5}. For $t = 0$, we obtain a GHZ state which has no NE, while for $t=1$, we have a W-state with a NE that coincides with the Pareto optimal choice. As we increase $t$ from $t=0$, we move away from the GHZ state and we see no NE up to $t=0.4$. There is no smooth transition of NE as we increase $t$ further. Near $t=1$ (W-states), we recover the NE found earlier (fig.\ \ref{fig:1}).

The quantum prisoners' dilemma has shed light on many aspects of quantum games including the existence of new NE not readily available in the classical games, and the absence of NE for maximally entangled inputs. The addition of entanglement to a third qubit controlled by the referee is seen to result in a rich structure that can be either beneficial, or detrimental to the players in terms of their payoff. If the bipartite entanglement between the players is near maximal, there is no NE, but by preparing the player's qubits near a W-state, the referee can steer the game so that it has a Pareto optimal NE. Interestingly, not all tripartite entangled states are equivalent, as the GHZ state gives no NE, unlike the W-state which yields NE that coincide with the Pareto optimal choice.


\end{document}